%% file: main.tex
\documentclass[final,5p,times,twocolumn]{elsarticle}

\usepackage{amssymb}
\usepackage{lipsum}
\usepackage{siunitx}
\DeclareSIUnit{\MeV}{MeV}
\DeclareSIUnit{\GeV}{GeV}
\usepackage[version=4]{mhchem}
\usepackage{svg}
\usepackage{textcomp}
\usepackage{lineno}
\usepackage{subcaption}
\usepackage{gensymb}
\usepackage{placeins}

\newcommand{\electron}{$\mathrm{e^-}$}
\usepackage{url}

\journal{Nuclear Physics A}

\begin{document}
\begin{frontmatter}

\title{Charge sharing and alignment performance of bent ALPIDEs measured with low-energy protons}

\author[a]{Berkin Ulukutlu\corref{cor1}}
\ead{berkin.ulukutlu@tum.de}
\cortext[cor1]{Corresponding author}
\author[a]{Christopher Ehrich}
\author[a]{Laura Fabbietti}
\author[a]{Roman Gernh\"auser}
\author[b]{Fabrizio Grosa}
\author[b]{Hartmut Hillemanns}
\author[a]{Tobias Jenegger}
\author[b]{Alex Kluge}
\author[a,b]{Lukas Lautner}
\author[b]{Magnus Mager}
\author[a]{Lukas Ponnath}
\author[c]{Andrea Rossi}
\author[a,b]{Isabella Sanna}
\author[d]{Serhiy Senyukov}
\author[e]{Johanna Stachel}
\author[b]{Miljenko \v{S}uljić}
\author[a]{Laszlo Varga}
\author[e]{Alperen Y\"unc\"u}

\address[a]{Technische Universit\"at M\"unchen, Munich, Germany}
\address[b]{European Organisation for Nuclear Research (CERN), Geneva, Switzerland}
\address[c]{INFN, Sezione di Padova, Padova, Italy}
\address[d]{Universit\'e de Strasbourg, CNRS, IPHC UMR 7178, Strasbourg, France}
\address[e]{Ruprecht Karls Universit\"at Heidelberg, Heidelberg, Germany}

\begin{abstract}
The upgrade of the ALICE experiments Inner Tracking System (ITS3) aims to replace its innermost detection layers with bent wafer-scale CMOS MAPS sensors. This study examines the performance of ALPIDE chips, currently used in the ALICE ITS2, when operated in a bent configuration under realistic experimental conditions. Proton beams with energies of \SIlist{80;120;200}{\MeV} were used to study proton-proton elastic scattering on a polypropylene fiber target reconstructed using two opposing arms of trackers with sensors bent to radii of \SIlist{18;24;30}{\milli\meter}. The measured low-momentum protons provided a testbed for investigating clustering behavior in high-energy loss events, where no significant impact of bending was observed on cluster size. Additionally, alignment strategies for bent detectors were evaluated using the distance of closest approach (DCA) and opening angle between scattered proton tracks as benchmarks. The achieved resolution matches expectations from simulations, confirming the suitability of bent MAPS sensors for future high-energy and nuclear physics applications.
\end{abstract}

\begin{keyword}
Monolithic active pixel sensors \sep Solid state detectors \sep Cylindrical detectors \sep Alignment 
\end{keyword}

\end{frontmatter}


\input{01_introduction}
\input{02_setup}
\input{03_cluster}
\input{04_alignment}
\input{05_conclusion}

\section*{Acknowledgements}
This project is supported by the Excellence Cluster ORIGINS funded by the Deutsche Forschungsgemeinschaft (DFG, German Research Foundation) under Germany's Excellence Strategy EXC-2094-390783311.

This work has been sponsored by the Wolfgang Gentner Programme of the
German Federal Ministry of Education and Research (grant no. 13E18CHA).

The authors would like to thank the personnel of the Cyclotron Center Bronowice (CCB), Poland, for providing excellent beam conditions and for their valuable technical support throughout the beam test measurements.


\bibliographystyle{elsarticle-num}
\bibliography{references}
\end{document}

%% file: 01_introduction.tex
\section{Introduction}
\label{introduction}
The ALICE (A Large Ion Collider Experiment) \cite{Colella_2022} detector at the Large Hadron Collider (LHC) \cite{Kuijer_2003} is aiming to upgrade its Inner Tracking System (ITS) \cite{Kluge_2022,The:2890181} to improve the tracking and vertexing capabilities of the experiment. The existing tracking system, ITS2, features seven concentric layers of Monolithic Active Pixel Sensors (MAPS) \cite{Yang_2015,alice_ls2_upgrade} and allows for an impact parameter (defined as the distance of closest approach between a primary track and the interaction point) resolution of  30 $\mu$m in both the radial and longitudinal directions for \SI{1}{\GeV}$/c$ transverse momentum pions \cite{Reidt_2021}. The upcoming ITS3 upgrade planned during the LHC Long Shutdown 3 (LS3) aims to push the boundaries of silicon detector technology further by replacing the innermost three layers, currently built of traditional flat sensors, with bent wafer-scale MAPS detectors \cite{Hillemanns_2023} thinned down to approximately \SI{50}{\micro\meter}, which can be bent into cylindrical shapes and wrapped around the beam pipe. By eliminating the need for extensive support structures and cooling systems, bent detectors can significantly reduce the per-layer material budget: the bent silicon sensor contributes only $\approx$0.07\% $X_0$ for ITS3, compared with $\approx$0.36\% $X_0$ for an ITS2 inner-barrel layer including its mechanical support and services. This material budget reduction, together with a closer positioning of the innermost layers and the reduced beam-pipe radius and thickness foreseen for the upgrade, is expected to improve the impact parameter resolution by up to a factor of two with respect to the ITS2 detector \cite{The:2890181}.

The reduction of the material budget is even more important in inverse kinematic heavy-ion experiments at lower energies, as in the R3B \cite{aumann2007r3b} experiment at FAIR \cite{belias2020fair}, Germany, or in the Samurai \cite{kobayashi2013samurai} experiment at RIKEN, Japan. The reconstruction of vertices and particle momenta in these experiments is typically limited by the position resolution and multiple scattering in the first layers of the recoil tracking detector. 
Replacing the currently used silicon strip detectors \cite{duer2022fourneutron}, first with thin monolithic pixel sensors and ultimately with bent wafer-scale MAPS, exhibiting a position resolution better than $\sigma_x < 10$~$\mu$m, will address both limitations at the same time. The extremely low material budget will not only reduce secondary reactions and background contributions, but will also significantly improve the reconstruction of particle trajectories, thereby enabling a new level of precision in particle four momentum reconstruction. First implementations of this technology are already on their way \cite{r3btrt}.     

The introduction of bent wafer-scale silicon detectors represents a significant technological leap. The impact of bending on the sensor performance and stability, both mechanical and electronic, needs to be studied. For this purpose, large-scale mechanical samples and functional MAPS devices, specifically ALPIDEs \cite{Mager_2016,alice_ls2_upgrade}, have been tested in a bent configuration. It was observed that \SI{50}{\micro\meter} thinned silicon devices can reliably be operated long-term with bending radii down to \SI{18}{\milli\meter} \cite{inbeam_bent_its}. Additionally, bending did not noticeably affect either the pixel performance or the operation of the readout periphery of the sensor \cite{The:2890181,Rossewij_2024}. Notably, piezoresistive effects in bent ALPIDE sensors can induce shifts in analog supply current and threshold voltages \cite{Rossewij_2024}, which could in principle affect charge collection uniformity across sensors at different bending radii. In the present study, individual threshold calibrations are performed for each sensor to ensure uniform operating conditions, as described in Section~\ref{clusterSize}.

In this study, we evaluate the performance of a detector consisting of five bent ALPIDE chips operated under realistic experimental conditions. The detector is set up in a fixed target configuration with the aim of reconstructing elastic (p,p) collisions exploiting proton beams of different energies (\SIlist{80;120;200}{\MeV}) and a solid polypropylene target.
The conducted experiment provides a testbed for two open questions directly relevant to ITS3: (i) whether bending alters the charge-sharing and cluster formation properties of MAPS sensors under high-energy-loss conditions — a regime not previously studied in bent configurations — and (ii) whether cylindrical detector geometries with limited track sampling can be aligned to the multiple-scattering limit without external reference detectors. Previous studies with bent structures were limited to minimum ionizing particle signals \cite{lukas2025bent}, leaving high-ionization performance uncharacterized. The extrapolation of these results to operational ITS3 conditions requires the assumption that bending-induced effects scale monotonically, though not necessarily linearly, from the tested radii of \SIrange{18}{30}{\milli\meter} to the ITS3 design radius of \SI{19}{\milli\meter}, and that the proton beam test conditions adequately probe the relevant charge transport properties of the sensor. No radius-dependent effects are observed across this range in the present data (Section~\ref{clusterSize}), supporting this assumption. For the R3B and Samurai programmes, the experiment directly emulates the tracking conditions of interest, with low-energy recoil protons traversing thin bent detectors.
Furthermore, alignment strategies for the cylindrical geometry, needed to maximize the impact parameter resolution, are tested and compared. The results of this study provide critical insights into the feasibility and benefits of using bent wafer-scale detectors in future high-energy and nuclear physics experiments.

%% file: 02_setup.tex
\section{Experimental Setup}
\label{setup}
  The experiment is conducted at the Cyclotron Center Bronowice, Poland, a medical facility equipped with a Proteus C-235 cyclotron \cite{IFJ_CCB}. The accelerator complex provides proton beams of variable energies to the experimental hall where the detector setup was located. The setup involves a small telescope featuring bent ALPIDE chips configured to measure scattered protons from a target array at three different beam energies of \SIlist{80;120;200}{\MeV}.

\begin{figure}
    \centering
    \includegraphics[width=1\linewidth]{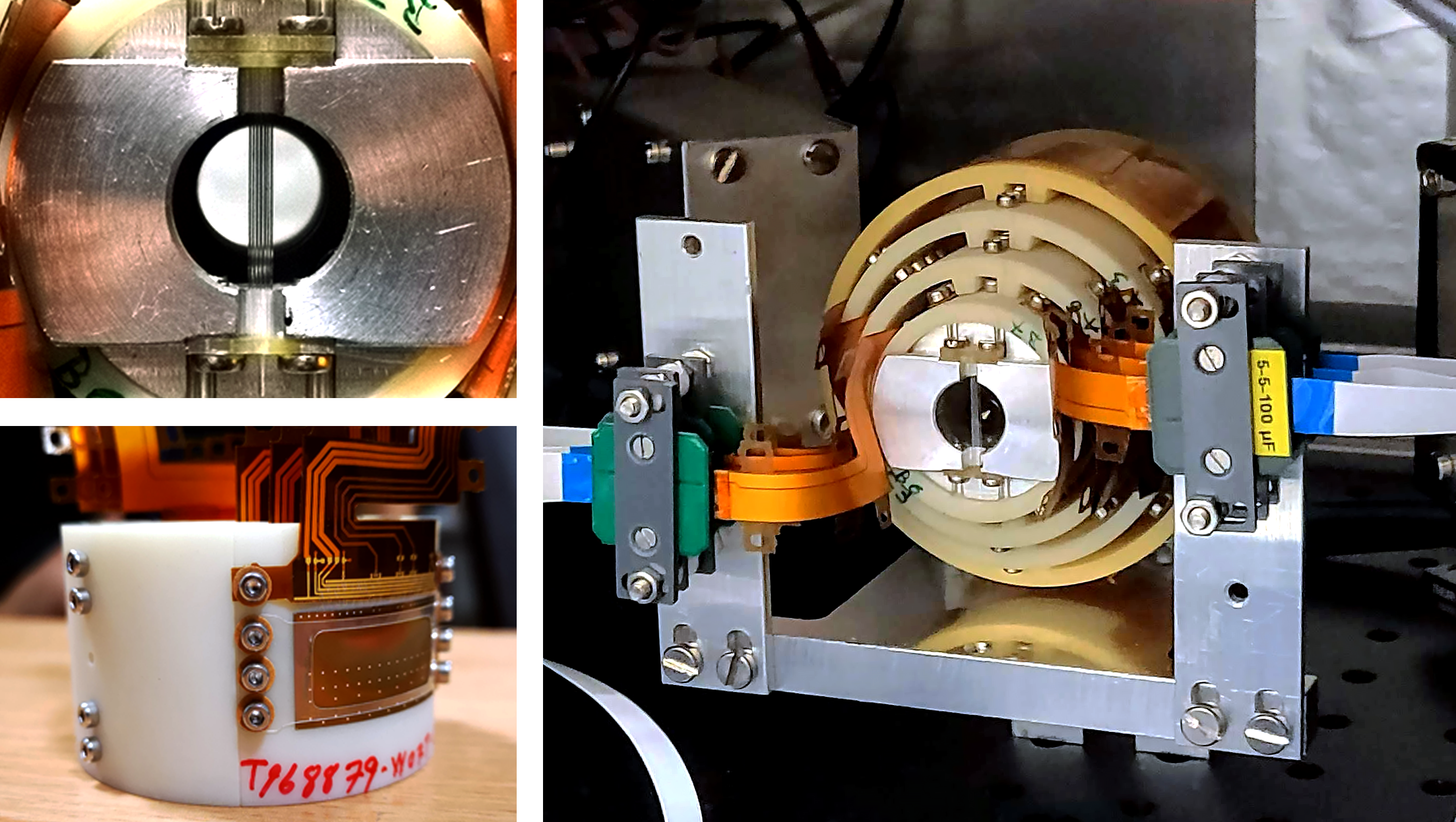}
    \caption{Photos of the detector setup. The used target fiber array (top left). Outer bent ALPIDE layer (bottom left). Full barrel installed in the beam setup (right).}
    \label{fig:photo}
\end{figure}

\subsection{Sensor}
The ALPIDE (ALICE Pixel Detector) is a state-of-the-art particle tracking sensor developed for the ALICE ITS2 upgrade using \SI{180}{\nano\meter} CMOS technology \cite{Rinella_2017,alice_ls2_upgrade}. The Monolithic Active Pixel Sensor (MAPS) architecture integrates the active sensor volume of the pixels, the analogue signal processing and the readout electronics on a single silicon substrate. The ALPIDE chips measure \SI{15}{\milli\meter} by \SI{30}{\milli\meter} with a matrix of 512×1024 pixels and a pixel pitch of 26.88×29.24 µm², achieving an intrinsic spatial resolution better than \SI{5}{\micro\meter} \cite{Reidt_2021,alice_ls2_upgrade}. The sensor has a low power consumption of \SI{47.5}{\milli\watt\per\centi\meter\squared} or \SI{35}{\milli\watt\per\centi\meter\squared}, depending on the operation mode \cite{alice_ls2_upgrade}. In ALICE ITS2, the ALPIDE sensors in the inner layers are thinned down to a thickness of \SI{50}{\micro\meter}; together with the lightweight mechanical support and cooling, this enables a very low material budget of 0.36\% $X_0$ in the inner layers \cite{Reidt_2021}. In this work, the ALPIDE sensors are used in a bent configuration (see Figs.~\ref{fig:photo} and \ref{fig:scattering}), made possible by the elasticity of thin silicon substrate. The available extensive characterisation of the sensor in the flat configuration \cite{alice_ls2_upgrade} is used as a baseline.

\subsection{Telescope configuration}
\begin{figure}
    \centering
    \includegraphics[width=1\linewidth]{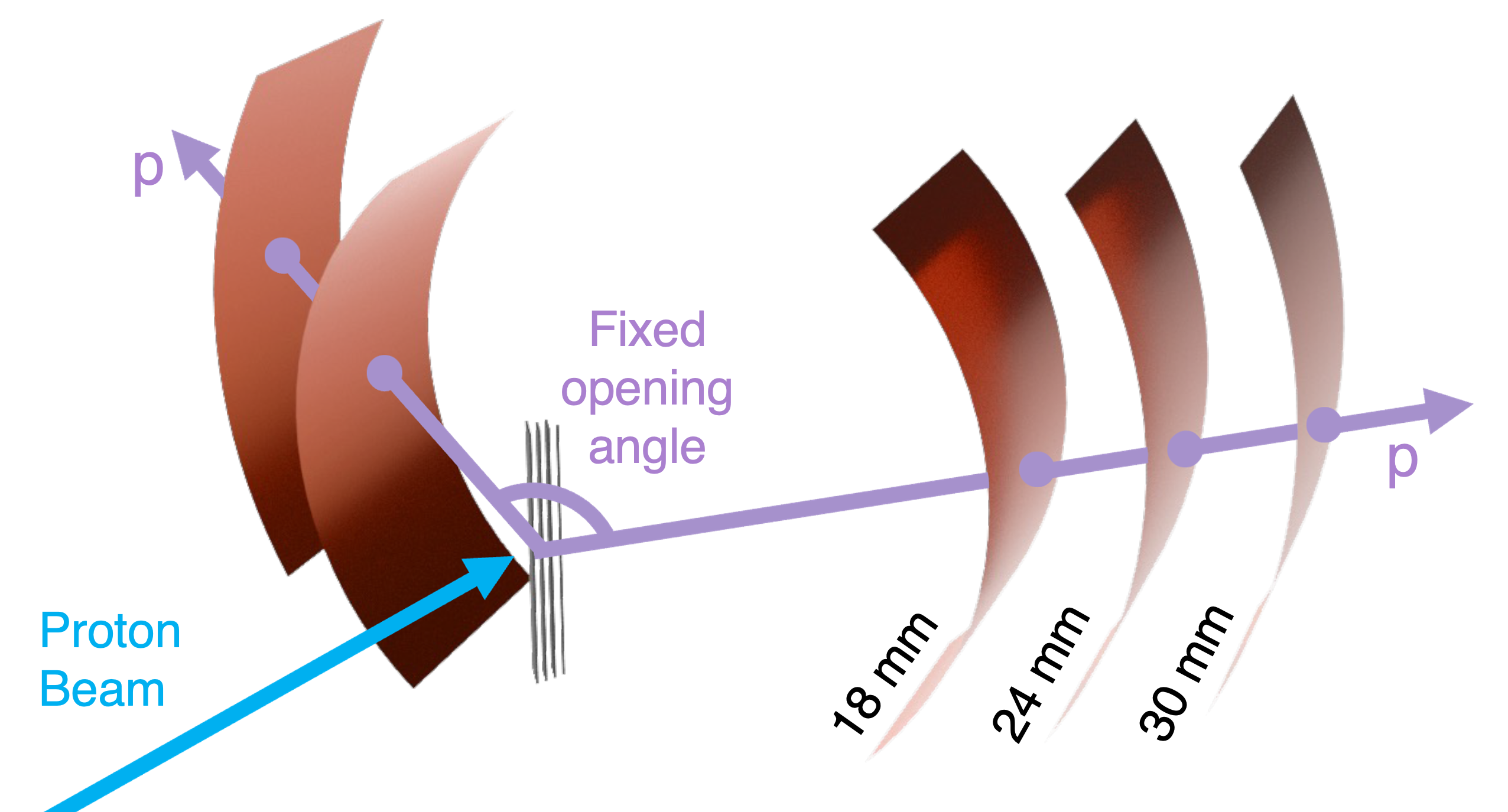}
    \caption{Sketch depicting an example event. The proton beams of various energies were shot on the target fiber array, where the outgoing elastically scattered protons were measured by the bent ALPIDE sensors.}
    \label{fig:scattering}
\end{figure}

The detector configuration featured ALPIDE sensors bent around three 3D printed cylindrical supports with radii of \SIlist{18;24;30}{\milli\meter}, arranged in a two-arm geometry pointing to the target (see Fig.~\ref{fig:scattering}). The telescope setup was intended to be operated with six sensors; however, due to damage to the middle layer on the left arm prior to data taking, the final setup features only five sensors — three on the right arm and two on the left arm. This loss of one of six planned sensors reduces the number of bending radii sampled on the left arm to two and eliminates the possibility of performing single-arm track residual minimization for alignment, as discussed further in Section~\ref{alignment}.
The sensors are assembled in the bent configuration by bending them onto 3D-printed mechanical jigs and holding them in place with a foil featuring a central cut‑out, as described in \cite{lukas2025bent}. The jigs feature a \SI{9}{\milli\meter} by \SI{17}{\milli\meter} opening window, which defines the Region of Interest (ROI) for the measured hits while the used radii emulate the cylindrically bent geometry planned for ITS3. 
On the long edge the ALPIDE sensors are wire-bonded to flexible printed circuit boards (FPCs), which interface the FPGA-based DAQ system \cite{daqBoard} using one readout board for each chip. The entire system is housed inside a metallic dark box (detector chamber) to minimize the influence of external light and electromagnetic interference.

The detector chamber is positioned in the experimental hall with its entrance window (\SI{12}{\micro\meter} of aluminum) \SI{40}{\centi\meter} downstream from the titanium vacuum window of the beam pipe. A beam spot diameter of \SI{4}{\milli\meter} (FWHM) was measured right at the target position of the telescope setup. The exact positioning of the chamber is tuned using a laser system such that the target is aligned with the beam.

The primary target comprises eleven polypropylene fibers (\ce{(C3H6)_n}) arranged in an arrow-like configuration pointing upstream, distributed in an equilateral hexagonal grid with a fiber-to-fiber separation of \SI{500}{\micro\meter}, as illustrated in Figs.~\ref{fig:photo} and~\ref{fig:scattering}. The geometric arrangement of the fibers is optimized to minimize secondary interactions of the scattered protons with adjacent fibers. Furthermore, the surrounding bent sensor layers are positioned to maximize acceptance for scattered protons emerging from the target region. The dominant reaction under investigation is the elastic scattering of protons on hydrogen nuclei within the fibers, which produces a characteristic pair of outgoing tracks separated by an opening angle of approximately \ang{90}, with relativistic corrections remaining below \ang{3} in the relevant energy range. In addition to the primary reaction channel, proton–carbon interactions can give rise to quasi-free scattering processes of the type $\mathrm{p(^{12}C,2p)^{11}B}$ \cite{jacob1966quasi}. Although this reaction channel is suppressed, it contributes to the background with a flat angular distribution of the two correlated protons without any pronounced peak structure. To evaluate the impact of this process on the measured angular distributions, dedicated simulations were performed. These simulations incorporated the internal Fermi motion of the bound proton, modeled with an isotropic momentum distribution characterized by a Gaussian width of \SI{105}{\MeV}$/c$ along each spatial axis, consistent with the nuclear Fermi momentum of $^{12}$C \cite{moniz1971nuclear}.

The sensors are triggered using two 3\,cm $\times$ 4\,cm $\times$ 5\,cm CsI(Tl) scintillators, developed within the CALIFA project \cite{cortina2014califa}, each positioned behind one of the tracker arms and thick enough to stop the scattered protons. The light signal is read out by one Hamamatsu Avalanche Photo Diode (APD) attached to the smallest surface of each crystal in downstream direction. The large volume scintillators have a larger acceptance compared to the sensors and ensure an efficient selection of elastic events by the coincidence pattern of two large signals, one in each arm.

\subsection{Data processing}
Data acquisition is managed using the EUDAQ2 software \cite{liu2019eudaq2}. Each recorded event (corresponding to a trigger signal from the scintillators) contains a timestamp and the row and column coordinates of all the pixels fired on the ALPIDE layers. Neighboring hits recorded in each event are later grouped into clusters using the Corryvreckan analysis framework \cite{dannheim2021corryvreckan}. The output from the Corryvreckan framework containing the measured cluster data is converted into a customized data format for further processing using Python \cite{uits3_krakow22} and ROOT \cite{root}. 
Dedicated software tools are developed to process the cluster data, enabling cluster hits to be selected according to the number of fired pixels to filter out background and noise-induced hits. Additionally, a track-finding algorithm is developed to reconstruct the event vertex, which is defined as the point of closest approach between the tracks detected in the two detector arms. Since the proton tracks of interest originate from singular points inside the target fibers, as shown in Fig.~\ref{fig:scattering}, beam-induced background — originating mostly from reactions of the beam with the exit window, the ambient air, and the holding structures close to the beam line — is removed by applying spatial selections to the reconstructed event vertices.  

%% file: 03_cluster.tex
\section{Cluster size and signal amplitude }
\label{clusterSize}
When a charged particle traverses the ALPIDE sensor, it can occur that more than one neighboring pixel registers a signal. This effect is mostly driven by charge diffusion in the not fully depleted epitaxial volume leading to the sharing of charge carriers between neighboring pixels. In the ALPIDE design \cite{Rinella_2017}, the depletion region does not extend to the pixel edges, which can further result in clusters with pixel counts larger than the number of next neighbors in large energy deposition hits. As such, the cluster size (the number of firing pixels making up a cluster) is related to the deposited energy and can be used for Particle Identification (PID). However, for employing such techniques in bent detectors, it is crucial to pin down possible effects due to the bending on the cluster properties. Previously, this question has been investigated with bent ALPIDEs measuring Minimum Ionizing Particles (MIP) signals in Ref. \cite{lukas2025bent}. 

In this experiment, low energy protons have an energy deposition which is on average 5 times larger than MIPs. This condition is ideal to study the influence of sensor bending on charge-sharing and clustering evolution.

\begin{figure}[ht]
    \centering
    \includegraphics[width=1\linewidth]{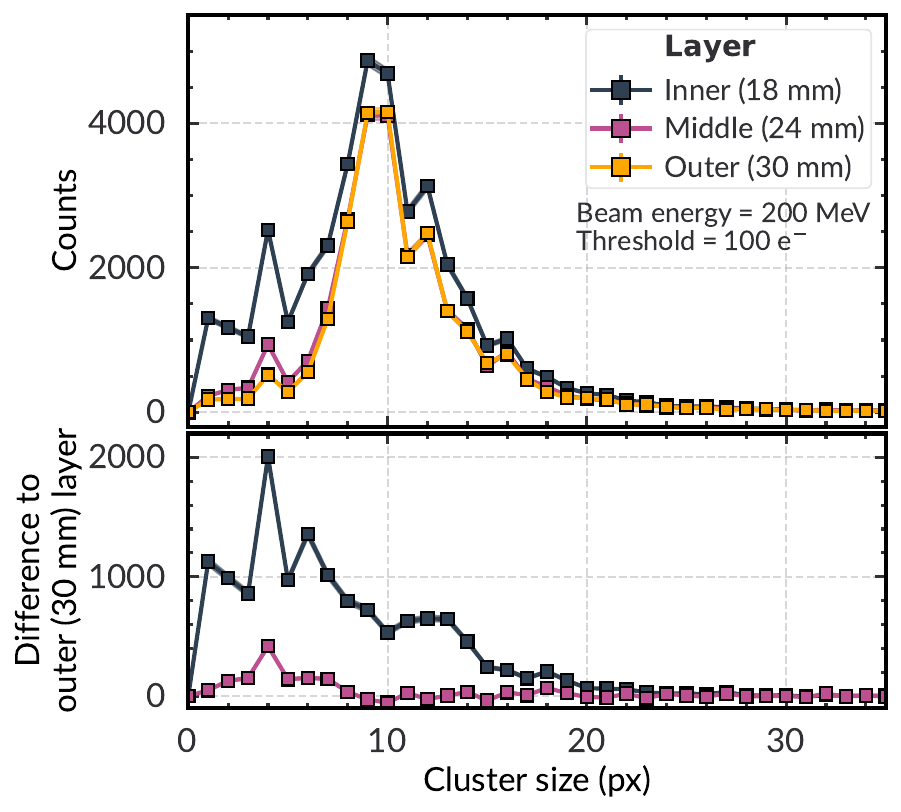}
    \caption{Raw cluster size distribution at \SI{200}{\MeV} proton energy and 100 $\mathrm{e^-}$ threshold value. An excess of broadly distributed clusters attributed to background from delta electrons can be observed in the inner layers. Cluster sizes corresponding to rectangular shapes (2x2, 3x3, 4x4) are visibly favored.}    
    \label{fig:All_Clusters_Plot}
\end{figure}

The measured raw cluster size distribution for all layers is shown in Fig.~\ref{fig:All_Clusters_Plot} for a beam energy setting of \SI{200}{\MeV} and the average pixel amplitude threshold value corresponding to about 100~\electron at a reverse substrate bias of $V_{\mathrm{bb}} = -1.2$~V (standard setting used in ALICE ITS2 \cite{isakov2025alice,its2_calibration}). 
The cluster distributions generally peak around 10 pixels, but the shapes vary across different layers. The inner layers, in particular, show a significantly larger contribution of smaller cluster sizes. This is predominantly attributed to low-energy secondary electrons — mostly delta electrons, with a smaller contribution from photon conversions — stopped in the first layers, as confirmed by Geant4 simulations, which lead to a wide energy deposition distribution on the inner layers (as shown in Fig.~\ref{fig:All_Clusters_Plot}). A subtraction of the cluster size distributions of the inner and outer layers, each normalized to account for the different geometrical acceptance of the layers, is presented in the bottom panel of Fig.~\ref{fig:All_Clusters_Plot} where the broad cluster size distribution of the mentioned background contributors can be observed.

Furthermore, it is recognized that some cluster size values, especially those correlated to rectangular shapes (2x2=4, 2x3=6), are dominant in this sample, while other values adjacent to them (3,5) appear to be suppressed. This is expected to be due to the symmetric charge diffusion among pixels. For larger energy depositions, this effect is similarly observed for cluster sizes of 3x3=9, 4x3=12, and 4x4=16.  
 
\subsection{Track angles and cluster sizes}
\begin{figure}[ht]
    \centering
    \includegraphics[width=1\linewidth]{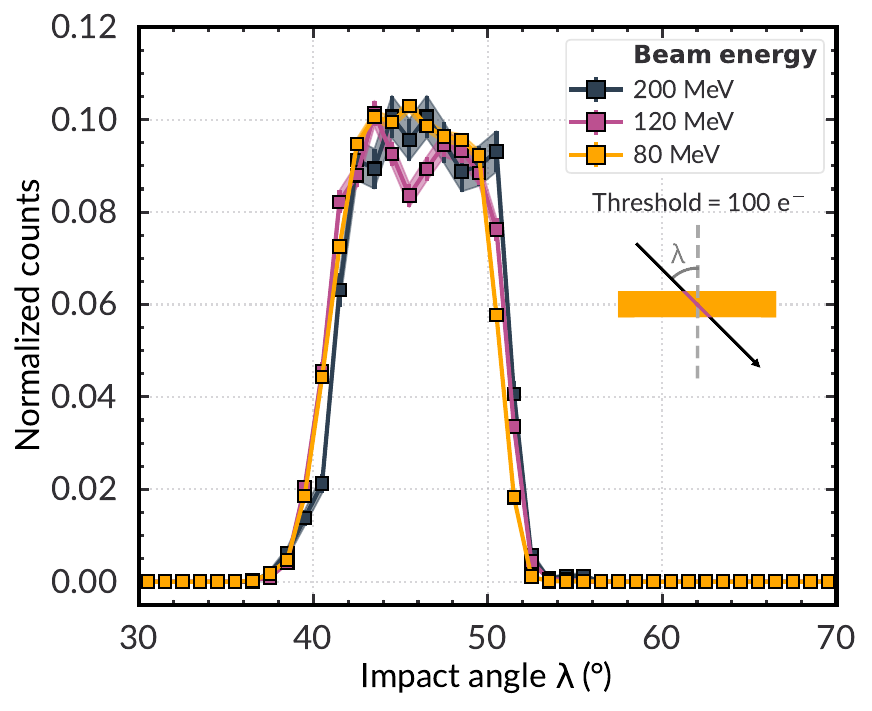}
    \caption{Distribution of impact angles between reconstructed tracks of interest and the detector layers.}
    \label{fig:impactAngleDistribution}
\end{figure}

The energy deposition from a track directly correlates with the path length the particles traverse in the active volume (effective thickness), which in turn is proportional to the incidence angle of the particles on the sensors. For the scattered proton tracks, the impact angle ($\mathrm{\lambda}$) is about \ang{45} as shown in Fig.~\ref{fig:impactAngleDistribution}. The impact angle $\lambda$ is defined as the angle between the fitted track, obtained from all three hits along the track, and the local normal to the cylindrical sensor surface at the hit point. The variation of this angle between layers due to multiple scattering is negligible, as the applied track-fit quality cuts reject strongly scattered tracks. The expected energy loss for the detected protons is, on average, a factor $\sqrt{2}$ larger compared to the expectation for perpendicular proton tracks with the same energy. To be able to use the cluster size information for particle identification of reconstructed tracks, the measured raw cluster sizes are normalized by $cos(\mathrm{\lambda})$. Track inclination enlarges clusters both geometrically, through the larger number of pixels crossed along the projected path, and through the increased energy deposition over the longer path length. For the studied impact angles of about \ang{45}, the geometrical term adds at most about two pixels to clusters exceeding ten pixels, so the enlargement is dominated by the energy deposition, which is removed by the $cos(\mathrm{\lambda})$ normalization.

\subsection{Cluster size and particle energy}
\begin{figure}[ht]
    \centering
    \includegraphics[width=1\linewidth]{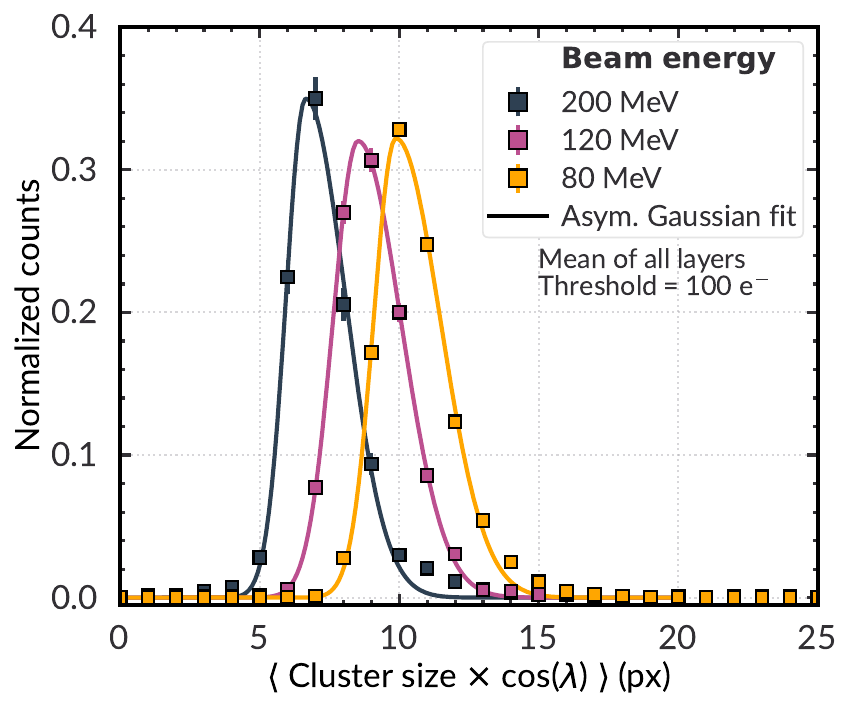}
    \caption{Average cluster size distribution from tracks associated to a vertex in the target region. With increasing beam energy the cluster size is observed to decrease, in agreement with the expected energy loss. The data points are fitted with an asymmetric Gaussian function to obtain the position of the most likely value.}
    \label{fig:clusterVsEnergy}
\end{figure}

As discussed before, cluster size depends mostly on two parameters: the amount of energy deposited in the detecting volume and the charge-collection threshold required to register a hit. Setting the latter constant, the cluster size of selected tracks originating at the target array, averaged over the three hits along each track measured on the three-layer arm, is shown as a distribution in Fig.~\ref{fig:clusterVsEnergy} for the three beam energies. The most probable value of this average-cluster-size distribution, extracted from an asymmetric Gaussian fit, is the quantity reported as a function of threshold and beam energy in Figs.~\ref{fig:threshold_and_energy_summarized} and~\ref{fig:clusterSizevsProtonMomentum}. Given the event kinematics and the narrow rapidity acceptance of the detector, the energy of individual protons can be approximated as half of the beam energy. As expected, the cluster signal decreases with increasing particle energy or, correspondingly, lower energy loss from protons in the detector material. 

\subsection{Threshold and cluster size}
\begin{figure}
    \centering    
    \includegraphics[width=1\linewidth]{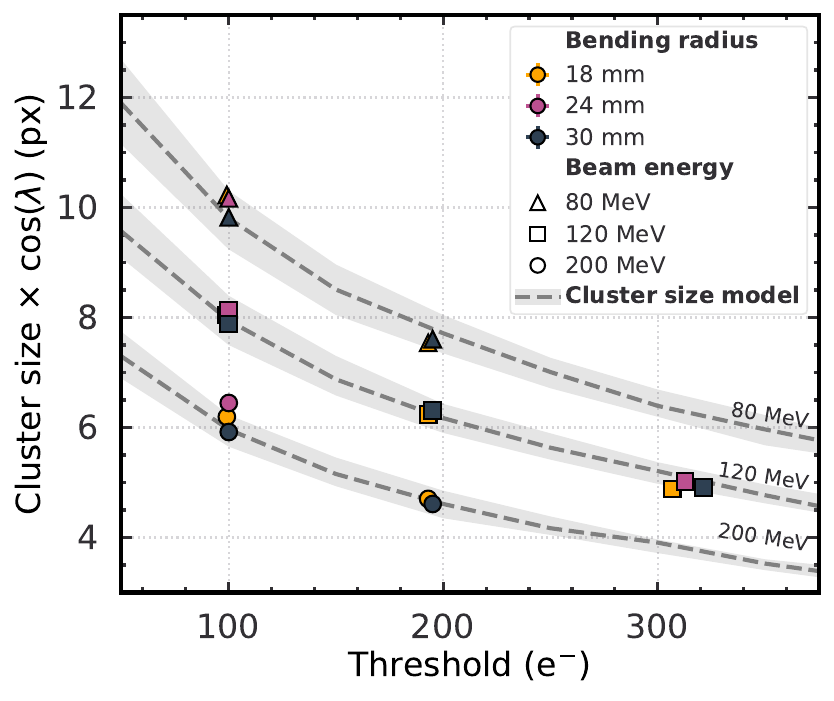}
    \caption{Scaling of average cluster size with the mean threshold set for various beam energies for different layers. The same trend is observed for all layers regardless of bending radius. The dashed gray line represents the trend from a phenomenological cluster size model assuming a 2D Gaussian charge diffusion profile, where the parameters of the profile, the width and scaling, are fitted to the data points. The bands represent the sensitivity of the parameterization to a 10\% variation of the fitted parameters around their best-fit values.}
    \label{fig:threshold_and_energy_summarized}
\end{figure}

Because the charge generated by a hit is shared among neighboring pixels, the charge-collection threshold directly determines how many pixels collect a charge above it and, thus, the measured cluster size. The relation between the threshold value and the measured cluster signal can be observed in Fig.~\ref{fig:threshold_and_energy_summarized}. The cluster size values are obtained by fitting the average-cluster-size distributions (shown as an example for a threshold value of 100 $\mathrm{e^-}$ and a beam energy of \SI{200}{\MeV} in Fig.~\ref{fig:clusterVsEnergy}) with asymmetric Gaussian functions, which enables the precise determination of the most probable value. The threshold is obtained by pulsing individual pixels on all sensors with varying amplitudes as described in \cite{its2_calibration}. The operational parameters of the detector layers were tuned to ensure the desired mean threshold values for all sensors.

A clear hierarchy between different sensors or different bending radii, respectively, was not observed in the conducted experiment. The layer-to-layer spread of most probable cluster size values is consistent with the statistical uncertainties of the individual fits, with no systematic ordering by bending radius. Within the sensitivity of this dataset — comprising sensors at radii of \SIlist{18;24;30}{\milli\meter} — a bending-induced shift in most probable cluster size exceeding approximately one pixel would have produced a statistically resolvable trend across the three radii. No such trend is observed. This observation from hits with high-energy depositions, combined with previously conducted studies using MIPs \cite{lukas2025bent} that similarly found no bending dependence, provides converging evidence that bending does not directly alter the clustering and charge sharing behavior of ALPIDE sensors across the tested radius range.

\subsection{Modeling cluster size in bent ALPIDEs}
A precise description of charge carrier diffusion between neighboring pixels requires detailed knowledge of the sensor’s electric field configuration and doping profiles. As an alternative, a phenomenological model can be constructed by assuming a symmetric two-dimensional Gaussian distribution for the spatial charge spread, enabling a practical parameterization of the observed cluster size trends as a function of deposited charge. The model parameters are fitted to the data and should be interpreted as a descriptive parameterization rather than a first-principles prediction.

In the model, the Gaussian profile is normalized such that its integral corresponds to the total deposited charge. The spatial extent of diffusion is characterized by the Gaussian width, $\sigma$, which directly controls the lateral spread of charge. The resulting continuous charge distribution is discretized according to the sensor's pixel pitch. For each pixel, the integrated charge is compared to a fixed threshold to determine whether it would trigger a binary response, consistent with the actual ALPIDE sensor behavior. The cluster size is then defined as the number of pixels with collected charge above this threshold.

To capture the observed dependence of diffusion on energy deposition, the $\sigma$ parameter is modeled with a power-law function: $\sigma_{x,y} = a\,Q_{\mathrm{total}}^{b}$, where $Q_{\mathrm{total}}$ is the total deposited charge, and $a$, $b$ are free parameters. These parameters are tuned to match the full data set featuring different beam energies (leading to different expected energy depositions) with multiple threshold settings as shown in Fig.~\ref{fig:threshold_and_energy_summarized}. Corresponding energy deposition values were obtained from Geant4 simulations (see Section~\ref{simulations}). The fit was performed using the Optuna hyperparameter optimization library in Python~\cite{optuna_2019}.

\begin{figure}
    \centering
    \includegraphics[width=1\linewidth]{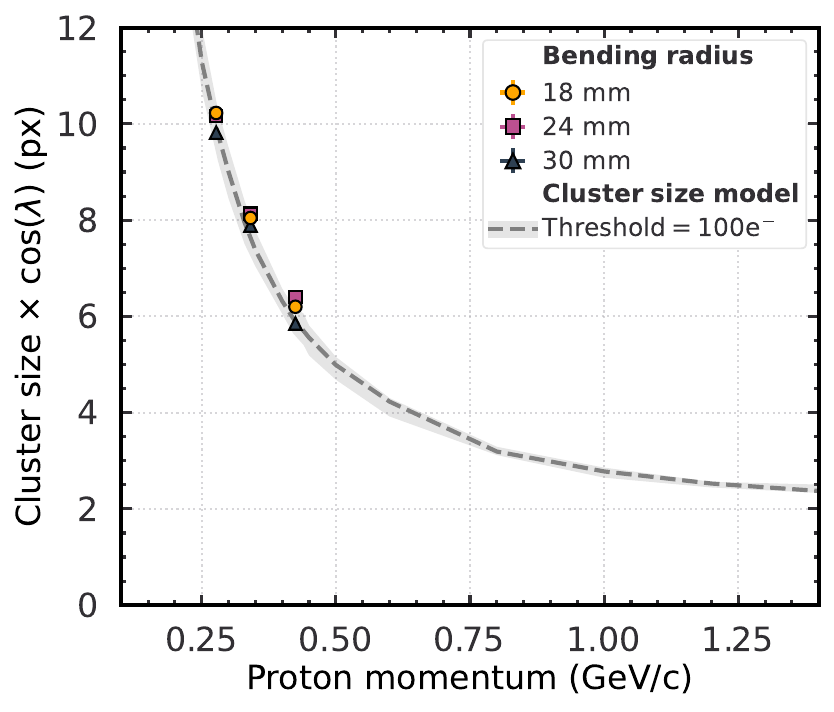}
    \caption{Average cluster size as a function of the proton momentum from the test beam data (colored symbols) compared to the phenomenological parameterization of cluster size scaling with energy loss (grey line with band). The parameterization is extended over a large proton momentum range, where it is observed to scale in agreement with the expected charge deposition from the Bethe-Bloch formula. The bands indicate the sensitivity to a 10\% variation of the fitted parameters around their best-fit values.}
    \label{fig:clusterSizevsProtonMomentum}
\end{figure}


Figure~\ref{fig:clusterSizevsProtonMomentum} shows the parameterized cluster size as a function of proton momentum for a threshold value of 100~$e^-$. The projected trend of the cluster size is in broad agreement with the expected charge deposition from Bethe-Bloch. Since the parameters are fitted to the same data used for this comparison, this agreement confirms internal consistency of the parameterization rather than constituting an independent validation. In addition to the observed absence of any bending-radius dependence in the cluster sizes, this result supports the potential use of the cluster size observable for particle identification in future bent detectors.

%% file: 04_alignment.tex
\section{Aligning bent detectors}
\label{alignment}
Accurate reconstruction of particle tracks from the position of the detector hits necessitates precise knowledge of the detector position. 
For traditional detector geometries, with sensors mounted onto planar holding structures in a flat configuration, different software tools exist to determine the optimal alignment corrections for the sensors. For bent detectors, the assumption of co-planarity of hits on the same detector layer does not hold. 

Another major challenge for aligning the test beam setup arises from the fact that one tracker arm has only two sensors, making it impossible to perform alignment via track residual minimization. Thus, the alignment must be carried out by optimizing global observables, such as minimizing vertex resolution and event plane residuals. Instead of modifying existing software libraries to accommodate the specific experimental layout featuring curved tracker layer surfaces, the development of various custom alignment algorithms is pursued. In this section, these alignment approaches are described and the performances are compared. 

The bending radius of the sensors is well constrained by the mechanical jigs onto which a polyimide cover foil presses down the sensors. Furthermore, measurements using an optical profilometer show that deviations from the expected radius in the region of interest for track reconstruction are limited to $\pm$\SI{20}{\micro\meter} \cite{lukas2025bent}. As a result, additional degrees of freedom, such as sagging and roll, are not significant for the experiment. Therefore, in this work, the corrections that were obtained are limited to translational and rotational degrees of freedom. However, sagging and roll are expected to become relevant for large areas and lightly supported bent detectors.

Using the mechanical precision of the setup, which was not especially optimized for the alignment of the three cylindrical detector surfaces, without any software alignment corrections, the RMS of the DCA distribution of the two tracks from both arms is \SI{835}{\micro\meter}. The RMS of the difference between the cluster position in the middle layer of the right arm and the intersection point of a straight line formed by the clusters in the other two layers is \SI{480}{\micro\meter} for the \SI{200}{\MeV} beam data. This is shown in Fig.~\ref{fig:event_plane_alignment} and indicates a significant systematic misalignment. 
Given the low energy of the protons being used in tracking and the low number of tracking layers (three and two in the two arms), the resolution in perfectly aligned cases is expected to be limited by multiple scattering. As a baseline for the comparison of the resolutions in the perfectly aligned detector configuration, Geant4 simulations are considered (see Section~\ref{simulations}).

\subsection{Kinematic constraints for pre-alignment}
\label{kinematic_constraints}
Elastic proton-proton scattering at low energies provides a clean two-body kinematics, where no excited states are involved.  
 The two protons are scattered back to back in the center-of-mass reference frame, meaning that they will be emitted from the target fiber and coplanar with the beam line. As a result, the azimuthal coordinates of the hits must be correlated not only along the same track, but also between the two tracks in an event.

\begin{figure}
    \centering
    \includegraphics[width=1\linewidth]{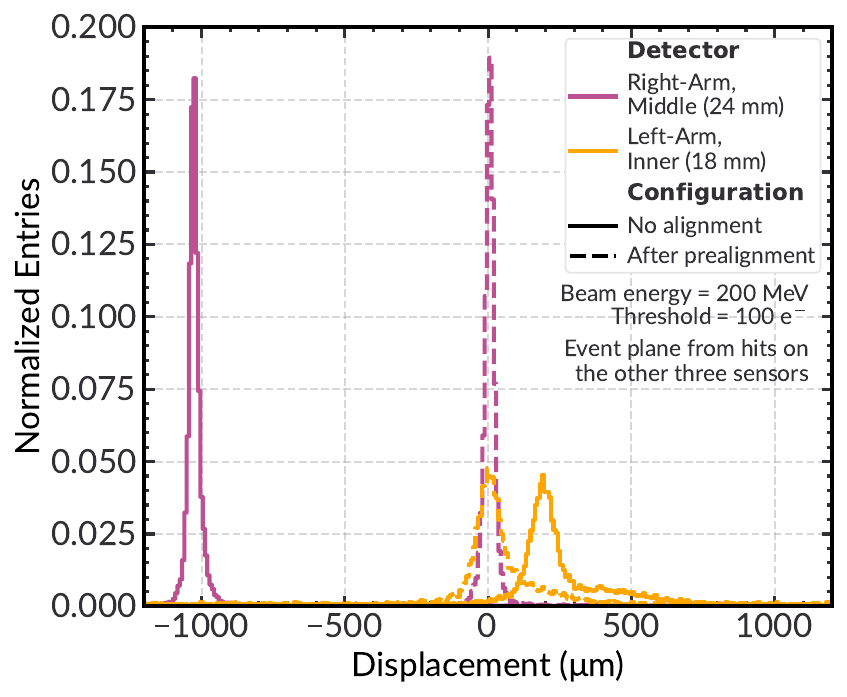}
    \caption{Example distributions of the residuals to the event plane for the different layers, before and after pre-alignment shown by the solid and dashed lines, respectively.}
    \label{fig:event_plane_alignment}
\end{figure}

This constraint is used to align the azimuth degree of freedom. Event planes are constructed, and the residuals of the hits recorded on different layers are calculated with respect to these planes. Given that any non-collinear three points can define a plane, for the alignment of each sensor, the plane is constructed using three other sensors. This procedure is carried out for each sensor, and corrections are obtained from the mean of the residual distribution, presented in Fig.~\ref{fig:event_plane_alignment}, where the distribution before and after the alignment is shown. It has to be observed that the middle layer sensor on the right arm is rotated by about \ang{2.5}. Correcting for this in the pre-alignment already results in a significant improvement of the performance as shown in Fig.~\ref{fig:event_plane_alignment}. Additional constraints are provided by the target geometry.

\subsection{Minimization alignment}
A more traditional approach is the Minimization alignment where the median two-track DCA, which depends on alignment quality, is minimized by exploring various geometric configurations. The two-track DCA is defined as the distance of closest approach between the two reconstructed proton tracks, which also defines the reconstructed vertex, and is hereafter also referred to as the vertex DCA. If the alignment is perfect, the DCA resolution should converge to the multiple scattering limit. The developed algorithm transforms the coordinates (azimuthal rotation and translation along the beam axis) of clusters from individual detectors and evaluates the DCA in search of better-performing configurations. The algorithm uses a bisection approach to avoid having to probe the entire space of possible detector positions. 

This algorithm relies on two assumptions. First, the minimal median vertex DCA should be a reliable proxy for the evaluation of the alignment. This assumption is sound, since the various positions of the target fibers forbid misalignment configurations where the two proton tracks are reliably focused into sharp vertex points. The fiber positions themselves are not imposed as a constraint in the alignment but are left free; only the existence of localized vertices in the target region is exploited. Second, while conducting the search over the detector configuration space, the evaluated misalignment should be convex. This is known not to be accurate when searching over only a small subset of degrees of freedom, where local minima arise. However, when all relevant degrees of freedom are considered, it should not be possible that a point more distant to the solution in alignment space yields better performance. With this, the minimization search was conducted by moving all sensors in all relevant translational and rotational degrees of freedom. As discussed at the beginning of this section, deviations of the bent sensors from the ideal cylindrical shape were measured to remain below \SI{20}{\micro\meter} over the region of interest \cite{lukas2025bent}; such deformations are therefore neglected, and only rigid rototranslational degrees of freedom are considered in the alignment.

Initially, the search range is defined for all degrees of freedom (parameters). At each step, the measured hit positions are corrected with the selected misalignment hypothesis by moving in 3D space accordingly. Afterwards, the tracks and vertices are reconstructed from the updated hits, and the alignment performance is evaluated. For the next step, the parameter configurations are selected using a bisection strategy, where for each parameter the mean of the previously tried pair is taken as the next one. For consecutive steps, search points are selected converging in the direction of the best performing known configuration. Since the search space is halved after each step, the precision desired (such as \SI{2}{\micro\meter} in our case) can be reached after a predetermined number of steps, depending on the starting parameter range. However, the total number of iterations grows exponentially with the number of degrees of freedom evaluated. To limit the computational cost, a small subset of events was used to determine the alignment, and the resulting misalignment correction was subsequently applied to the entire dataset.

\begin{figure}
    \centering
    \includegraphics[width=1\linewidth]{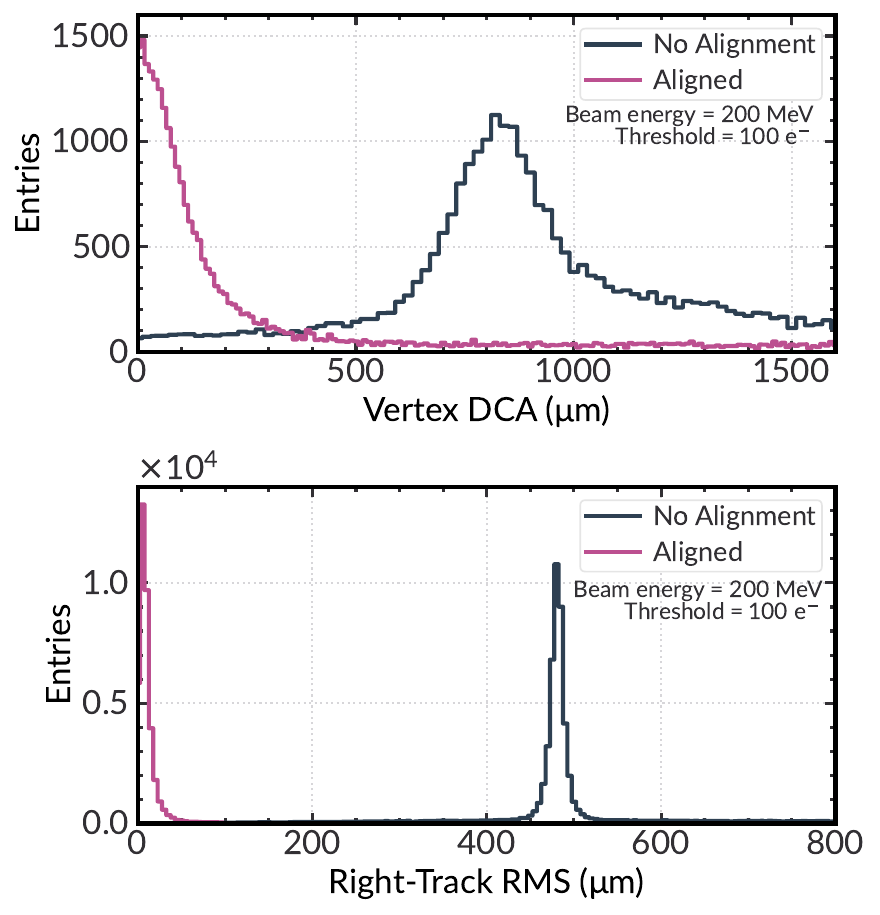}
    \caption{Reconstructed vertex DCA resolution (top) and the RMS of the residual between the reconstructed cluster and interpolated track position in the middle layer of the right-arm (bottom) before and after pre-alignment for the \SI{200}{\MeV} beam energy dataset.}
    \label{fig:DCA}
\end{figure}

\begin{figure*}
    \centering
    \includegraphics[width=1\textwidth]{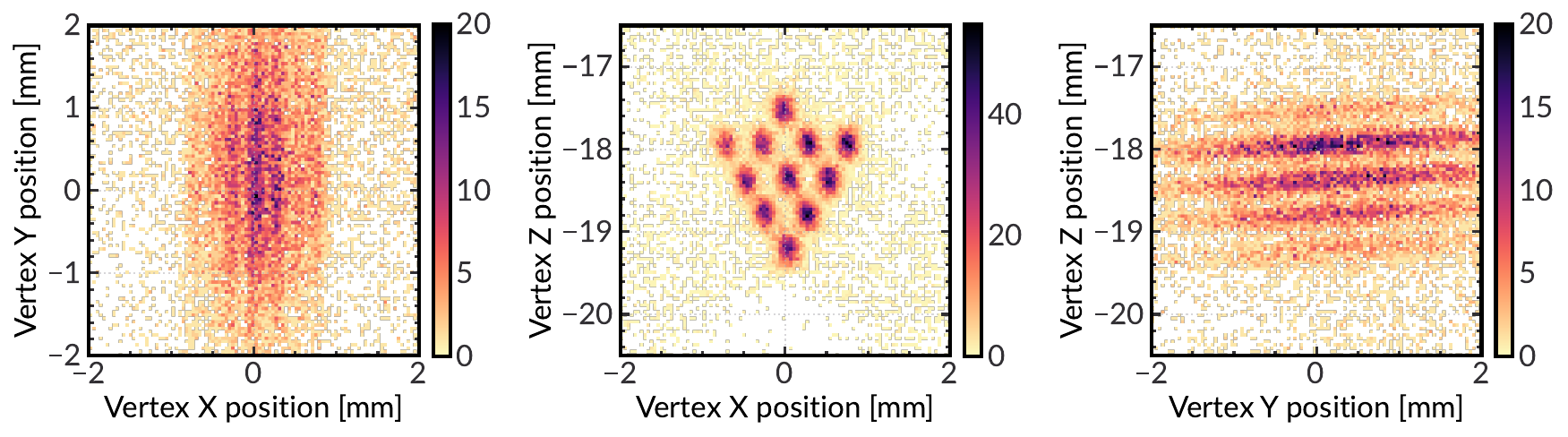}
    \caption{The projections of the vertex distributions after alignment, (left) front-view, (middle) top-view, (right) side-view of the target array.}
    \label{fig:vertex_distributions}
\end{figure*}

The improvement in the performance of the obtained alignment is shown in Fig.~\ref{fig:DCA}. The approximately flat tail visible in the DCA distribution (Fig.~\ref{fig:DCA}, top) is attributed to residual beam-induced and combinatorial background surviving the vertex selection. The projections of the reconstructed vertex distribution after alignment are shown in Fig.~\ref{fig:vertex_distributions}, where the individual target fibers are resolved in the front, top, and side views of the target array. Using the stricter selections described in Section~\ref{setup}, the resulting vertex DCA performance is presented in Fig.~\ref{fig:dca_data_vs_geant}, where a vertex DCA width of $(80.5 \pm 4.6)$~\si{\micro\meter} is achieved for the \SI{200}{\MeV} beam data. The optimization was terminated once the applied correction step size fell below \SI{1}{\micro\meter}.

\subsection{Performance and Comparison to GEANT4 Simulations}
\label{simulations}
\begin{figure}
    \centering
    \includegraphics[width=1\linewidth]{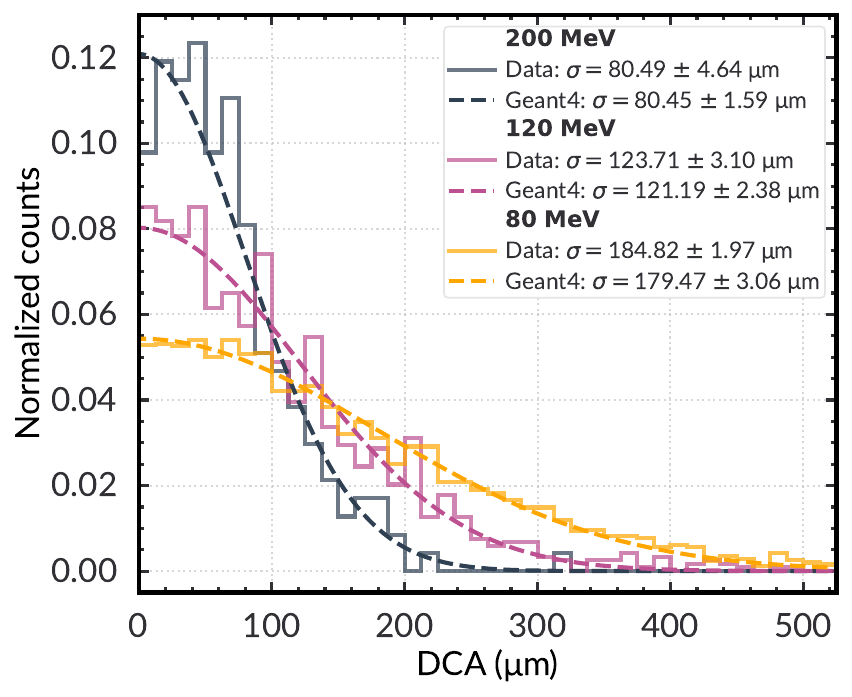}
    \caption{DCA distributions obtained for the final alignment, compared to expected distribution from Geant4 obtained including multiple scattering.}
    \label{fig:dca_data_vs_geant}
\end{figure}

Simulated data, as shown in Fig.~\ref{fig:dca_data_vs_geant}, are produced for various beam energies using Geant4. These simulations predict the optimal achievable vertexing resolution, considering multiple scattering effects in an ideally aligned configuration. For a correct comparison, the material dimensions and distances identical to the real detector within the region-of-interest are implemented. The simulation involves proton projectiles hitting the fiber target array, with only events featuring elastic proton-proton collisions within a target fiber being considered in the analysis. The detector layers are described by cylinders of silicon with a thickness of \SI{50}{\micro\meter}. The hit locations on the detector layers for the outgoing protons, as well as the trajectory of the protons are stored for the analysis. The recorded hit positions are smeared according to the expected sensor spatial resolution. Since the hit position is determined by the mean position of the pixels making up a cluster, the spatial resolution is improved compared to the intrinsic resolution available from a single pixel. In the conducted experiment, the cluster size is shown to depend on the beam energy, as such the smearing is applied on the simulated hits also reflecting this fact, by being sampled from a normal distribution with a standard deviation given by
\begin{equation}
    \sigma_{x,y} = \frac{\text{Pixel pitch}}{\sqrt{\langle \text{Cluster size} \rangle_{x,y} \times 12}}
\end{equation}
for each beam energy.

\begin{figure}
    \centering
    \includegraphics[width=1\linewidth]{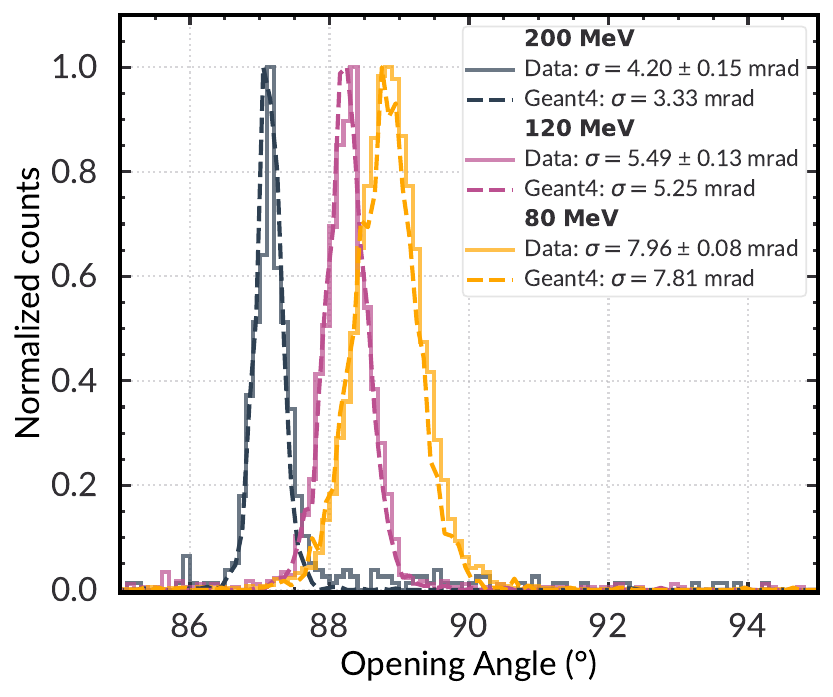}
    \caption{Opening angle from selected elastic scattering events compared to expectation from Geant4.}
    \label{fig:opening_angle}
\end{figure}

The comparison between the best obtained alignment from the described algorithm and the expectations for a perfectly aligned geometry simulated in Geant4 is presented in Fig.~\ref{fig:dca_data_vs_geant}. A good agreement between the simulation and the data can be observed. The measured DCA width, $(80.5 \pm 4.6)$~\si{\micro\meter}, agrees with the Geant4 multiple-scattering limit, $(80.5 \pm 1.6)$~\si{\micro\meter}: their difference of \SI{0.04}{\micro\meter} is far smaller than the combined statistical uncertainty of \SI{4.9}{\micro\meter}, which bounds any residual misalignment contribution to the DCA resolution at \SI{200}{\MeV}. The achieved resolution is therefore predominantly determined by multiple scattering rather than residual misalignment. It is noted that within the experimental limitations of the available beam energies and statistics, the intrinsic spatial resolution of the ALPIDE sensors does not contribute measurably to the DCA, and the alignment cannot be independently verified against that limit.
The agreement between the data and the simulation for the opening angle and the resolution of the angle distributions is presented in Fig.~\ref{fig:opening_angle}. The measured mean opening angles of \ang{87.1}, \ang{88.3}, and \ang{88.9} at \SIlist{200;120;80}{\MeV}, respectively, match the simulated expectations to within \SI{1}{\milli\radian}. Furthermore, it is demonstrated that the low material budget of the bent detector layers in the region of interest leads to the expected angular precision.

%% file: 05_conclusion.tex
\section{Summary and conclusions}
\label{conclusion}
In this work, a proton scattering experiment was carried out using bent ALPIDE sensors. The response of the bent sensors to low-energy scattered protons (down to \SI{40}{\MeV}) was investigated through measurements of cluster sizes, while tracking performance was studied using a dedicated alignment approach.

No statistically significant dependence of cluster size or charge sharing on bending radius was observed across sensors bent to radii of \SIlist{18;24;30}{\milli\meter}, at proton beam energies of \SIlist{80;120;200}{\MeV}. Within the sensitivity of this measurement — limited to about one pixel in the most probable cluster size by the resolution and the other contributions to the cluster-size spread — no bending-induced shift is observed, and the clustering behavior in bent silicon follows the expected systematics from prior measurements with minimum ionizing particles on planar samples \cite{lukas2025bent}. These two independent measurements, spanning the MIP regime and energy depositions up to ten times larger, provide converging evidence that charge sharing is not significantly altered by bending of the sensor across the tested radius range. The threshold dependence of cluster size was confirmed to follow the expected trend: increasing the charge collection threshold reduces the cluster size, with the same behavior observed at all bending radii, indicating that the threshold dependence of the cluster size is unaffected by the bent sensor geometry. The cluster size scaling with energy deposition is described by a phenomenological model assuming a symmetric two-dimensional Gaussian profile of the charge spread, fitted to the data, and is consistent with the Bethe-Bloch expectation.

The bent geometry, combined with the small number of tracking layers per arm (three and two on the right and left arms, respectively), poses a significant challenge for alignment. No standard track-residual-based alignment was possible for the two-sensor arm. Two custom alignment strategies were developed and compared. Initial misalignment was reduced using a pre-alignment approach exploiting the kinematic constraints of elastic proton-proton scattering; the coplanarity of the two outgoing tracks with the beam axis provides an event-plane constraint used to correct the azimuthal degrees of freedom. The best alignment was obtained using a bisection-based minimization algorithm, which iteratively converges on the optimal sensor configuration by minimizing the median DCA. Using this approach, a DCA width of $(80.5 \pm 4.6)$~\si{\micro\meter} was obtained at \SI{200}{\MeV}, in agreement with the Geant4 multiple-scattering limit of $(80.5 \pm 1.6)$~\si{\micro\meter}; their difference of \SI{0.04}{\micro\meter} is well within the combined statistical uncertainty of \SI{4.9}{\micro\meter}, which bounds any residual misalignment contribution to the DCA.

This experiment demonstrates the feasibility of operating cylindrically bent MAPS sensors in realistic in-beam conditions and aligning them to the multiple-scattering limit without external reference detectors. The results further support the viability of using bent wafer-scale CMOS MAPS sensors in the low-material-budget tracker layers planned for ITS3 and future nuclear physics experiments.